\documentclass[a4paper,11pt]{article}
\usepackage{pos}
\usepackage{pgfplots}
\usepackage{relsize}
\usepackage{slashed}
\usepackage{wrapfig}
\usepackage{graphicx} 
\usepackage{subcaption}
\renewcommand{\logo}{\relax}

\newcommand{\SModelS}{{\tt SModelS}~2.3.3}

\title{Light new physics in the top quark sample from the Large Hadron Collider}

\author*[a]{Dibyashree Sengupta}

\affiliation[a]{INFN, Laboratori Nazionali di Frascati, Via E. Fermi 54, 00044 Frascati (RM), Italy}

\emailAdd{Dibyashree.Sengupta@lnf.infn.it}

\abstract{Contrary to the general trend of looking for new physics at energies beyond the current reach of the Large Hadron Collider (LHC), this article proposes a strategy to look for light new physics via a meticulous study of well known and well-measured kinematic distributions. In this article, we propose performing such a study in the top-quark sample since the LHC, being a top-quark factory, helps in precise measurement of several observables related to the properties of the top-quark. One such observable is the invariant mass $m_{b\ell}$ of the b-jet and the charged lepton obtained from fully leptonic decay of pair-produced $t \bar{t}$ events. Such a strategy can be employed to extract hints for any Beyond Standard Model (BSM) scenario that allow for an exotic particle with mass close to the mass of top-quark ($m_t$) and can yield the same final state as fully leptonic decay of pair-produced top quarks. To provide a concrete study, we analyze a supersymmetric scenario with light right-handed stop quark with mass $\approx m_t$. The particle spectrum is such that the mass differences between the particles involved in the signal are small enough to lie in a potential blindspot and may be not yet firmly excluded by current LHC searches. Such spectra can yield a deviation from the Standard Model prediction in the lower region of the $m_{b\ell}$ distribution. This feature can be observed in any BSM framework that harbours light new physics that have so far escaped the LHC searches and hence can be used to extract light new physics signal irrespective of the underlying theory.}

\FullConference{42nd International Conference on High Energy Physics (ICHEP2024)\\
18-24 July 2024\\
Prague, Czech Republic\\}

\begin{document}
\maketitle

\section{Introduction}
\begin{figure} [h!]
    \centering
    \includegraphics[width=10cm]{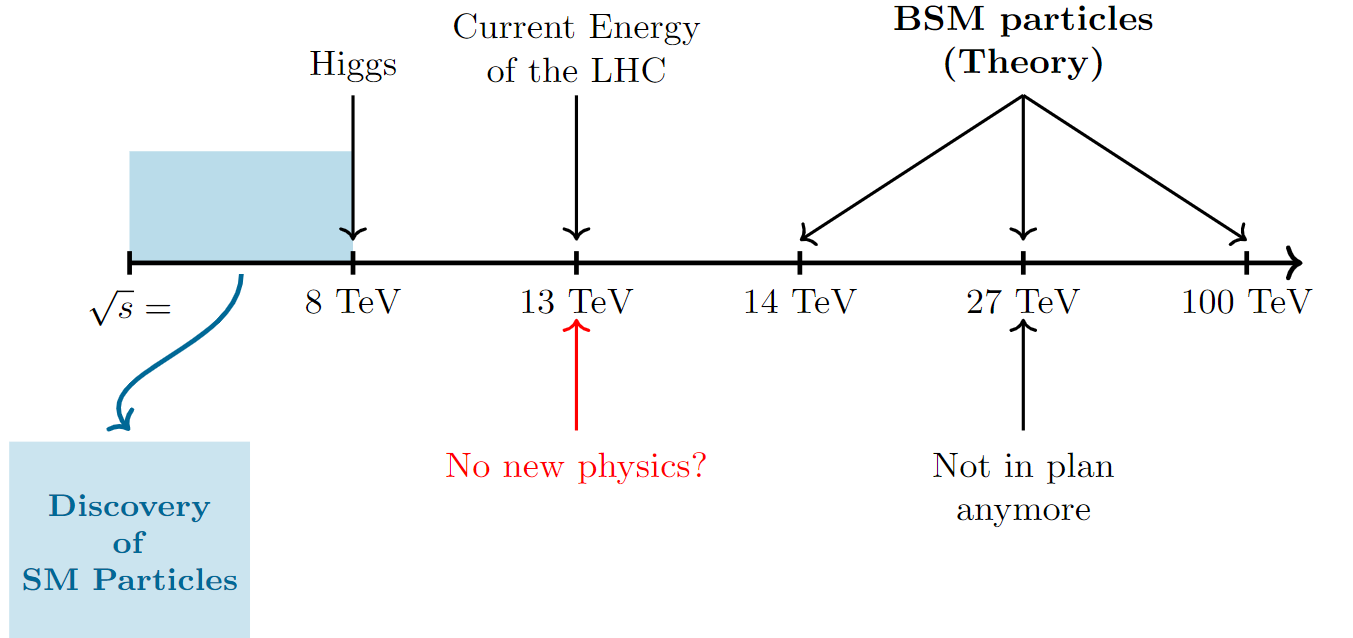}
    \caption{Energy line of SM and BSM particles.}
    \label{fig:energy_line}
\end{figure}

The Standard Model (SM), till today, is the most celebrated theory of nature. With the discovery of SM-like Higgs~\cite{Aad:2012tfa, Chatrchyan:2012ufa} at ATLAS and CMS in 2012, the particle spectrum of the SM is completed. Despite its tremendous success, the SM is not the complete theory of nature as it cannot explain several natural phenomena. Therefore, it is essential to introduce Beyond Standard Model (BSM) scenarios which can address one or more of the above issues. However, as depicted by Fig.~\ref{fig:energy_line}, even after a decade of the discovery of the SM-like Higgs, experiments have not found any hints for new particles that can vouch for any of the existing well-motivated BSM scenarios. Most of the existing phenomenological studies of several BSM scenarios~\cite{Bose:2022obr, CidVidal:2018eel} predict the discovery of new physics at colliders with energies and luminosities beyond the current reach of the Large Hadron Collider (LHC). However, the goal of this work is to propose a method to extract hints of new physics by precise studies of already well-measured and well-studied observables. In this work, we particularly focus on precise study of the $m_{b\ell}$ distribution in the top quark sample. We choose the top quark sample because the LHC is a top quark factory and hence assists in precise measurement of observables related to features of the top quark. We were able to devise a model-independent methodology to look for light new physics when precise measurement of top quark observables deviate from SM prediction, thereby, opening a plethora of possibilities of new strategies for searching new physics. Further details on this work can be found in Ref.~\cite{Bagnaschi:2023cxg}. With this understanding, the rest of the article is organized as follows. In Sec.~\ref{sec:tt}, we discuss how one can get hints of new physics in low energy regime, within the access of the LHC at present, from the top quark sample followed by a methodology for extracting this new physics signal in Sec.~\ref{sec:method}. Finally, in Sec.~\ref{sec:con}, we conclude. 

\section{Light new physics from $t \bar{t}$}
\label{sec:tt}

\begin{figure} [h!]
\begin{subfigure}[h]{0.32\linewidth}
\includegraphics[width=\linewidth]{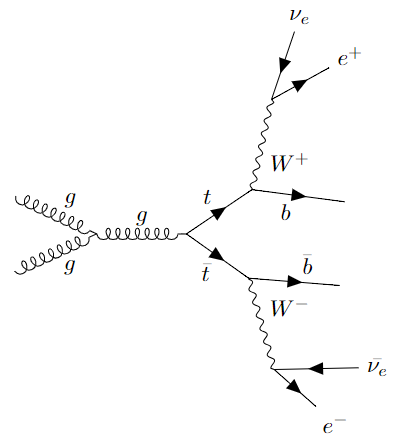}
\caption{SM Background}
\label{bma}
\end{subfigure}
\hfill
\begin{subfigure}[h]{0.3\linewidth}
\includegraphics[width=\linewidth]{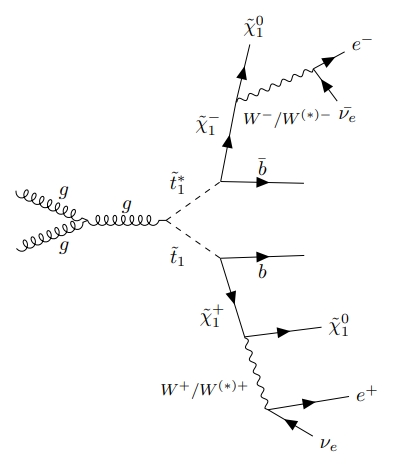}
\caption{MSSM Signal}
\label{bmb}
\end{subfigure}
\hfill
\begin{subfigure}[h]{0.3\linewidth}
\includegraphics[width=\linewidth]{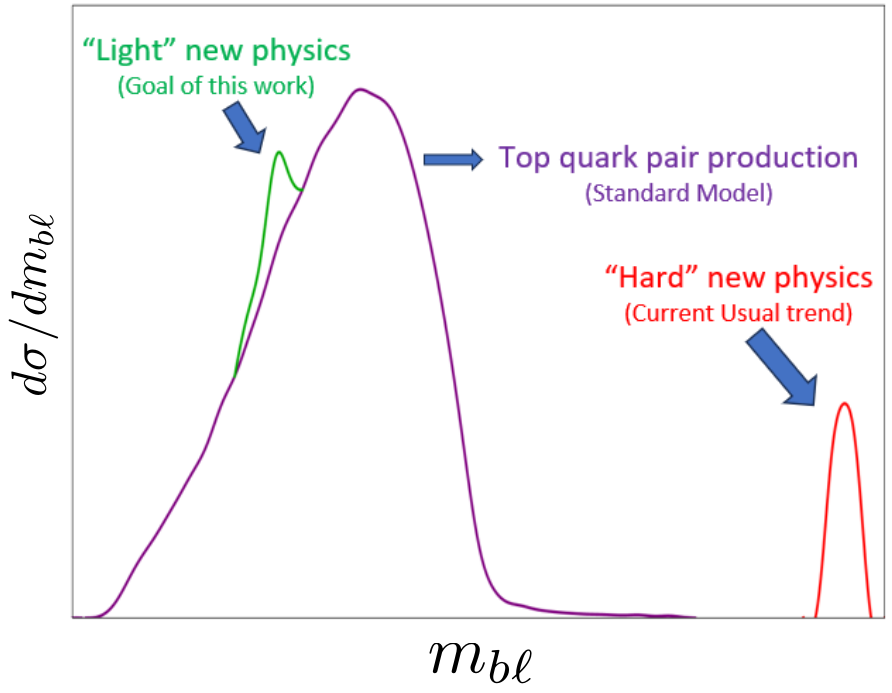}
\caption{$m_{b \ell}$ distribution}
\label{bmc}
\end{subfigure}%
\caption{}
\label{fig:bm}
\end{figure}

There have been several searches for new physics in the top quark sector ~\cite{Franceschini:2023nlp}. Generally, such searches involve large energy release owing to huge mass difference among the new exotic particles themselves and between these new exotic particles and SM particles. However, when such mass differences in the spectrum is small, such search strategies hit a blindspot. This issue of blindspot had appeared long ago in the LHC data ~\cite{Fan:2015mxp}. Probable solutions involved precise measurement of $t \bar{t}$ production cross-section ~\cite{Eifert:2014kea, Cohen:2019ycc}, deviation from SM prediction in measurement of mass of top quark $m_t$ ~\cite{Cohen:2019ycc} and angular distributions ~\cite{Cohen:2018arg, Han:2012fw} and kinematic distributions ~\cite{Macaluso:2015wja} of high-$p_T$ top quarks which might be sensitive to new physics. 
In spite of such solutions, one has not yet found any affirmative result yet in this regard. Therefore, in this article, we propose a model-independent method to look for new physics in the top quark sector. This method can be applied to extract any BSM scenario that can accommodate an exotic particle with mass close to the top quark and give rise to a final state similar to top quark pair production followed by their leptonic decay as shown in Fig.~\ref{bma}. As an example, we chose the Minimal Supersymmetric Standard Model (MSSM) scenario, an extremely well-motivated BSM scenario, with mass of the lightest stop quark $m_{\tilde{t}_1}$ around $m_t$. The signature of interest is pair-production of $\tilde{t}_1$ followed by their decay to $b-$ jet and the lightest chargino $\tilde{\chi}_1^{\pm}$ which further decays leptonically to the lightest neutralino $\tilde{\chi}_1^{0}$ which is also the lightest supersymmetric particle (LSP) in this framework. The relevant Feynman diagram is shown in Fig.~\ref{bmb}. Since we consider R-parity conserving MSSM scenario, the LSP forms the dark matter candidate and hence appears as large $\slashed{E_T}$  at the LHC. Therefore, the final state is composed of 2 $b-$ jets, a pair of oppositely-charged leptons and $\slashed{E_T}$. We propose that a closer look at the distribution of the invariant mass of a $b-$ jet and the lepton, denoted by $m_{b \ell}$, will show some deviation from the SM prediction in lower value of $m_{b \ell}$ to reveal light new physics. The purple curve in Fig.~\ref{bmc} depicts the $m_{b \ell}$ distribution expected from leptonic decay of pair-produced top quarks in the SM framework. The usual trend is to look for ``hard" new physics that will appear as a peak in the higher values of $m_{b \ell}$, as denoted by the red curve in Fig.~\ref{bmc}. However, the goal of this article is to look for light new physics which will appear in the lower $m_{b \ell}$ region as deviation from SM prediction, denoted by the green curve in Fig.~\ref{bmc}.   

\section{Methodology}
\label{sec:method}
\begin{figure}[h!] 
  \begin{subfigure}[h]{0.5\linewidth}
    \centering
    \includegraphics[width=.7\linewidth]{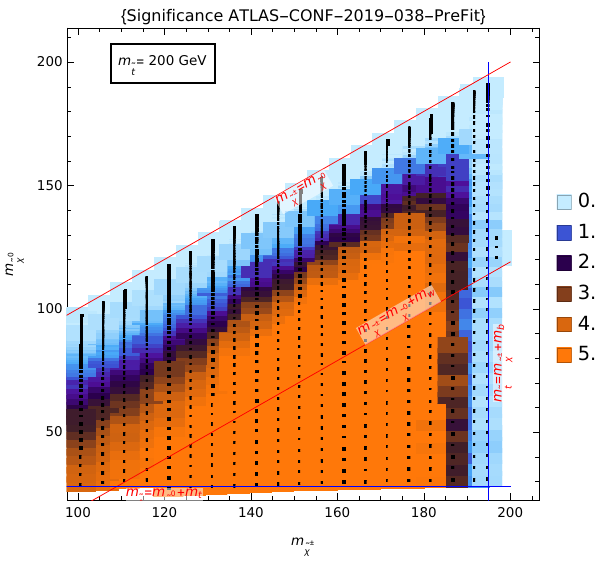} 
    \caption{Significance}
    \label{a}
  \end{subfigure}
  \begin{subfigure}[h]{0.5\linewidth}
    \centering
    \includegraphics[width=.7\linewidth]{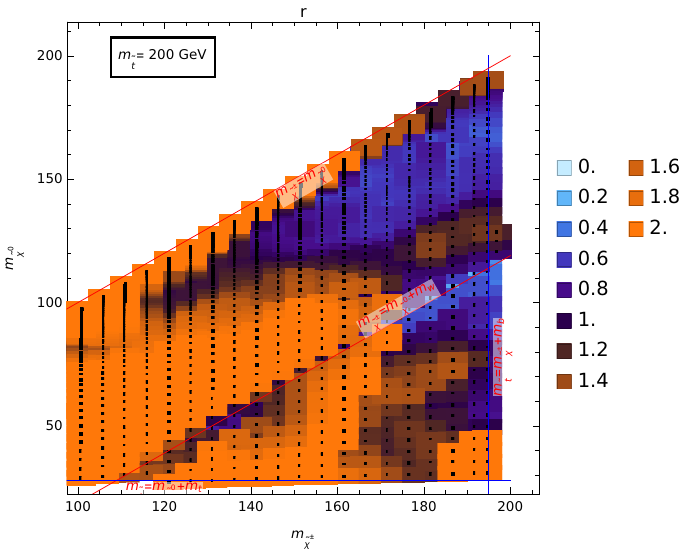}  
    \caption{$r$ from \SModelS}
    \label{b}
  \end{subfigure} 
  \begin{subfigure}[h]{0.5\linewidth}
    \centering
    \includegraphics[width=.7\linewidth]{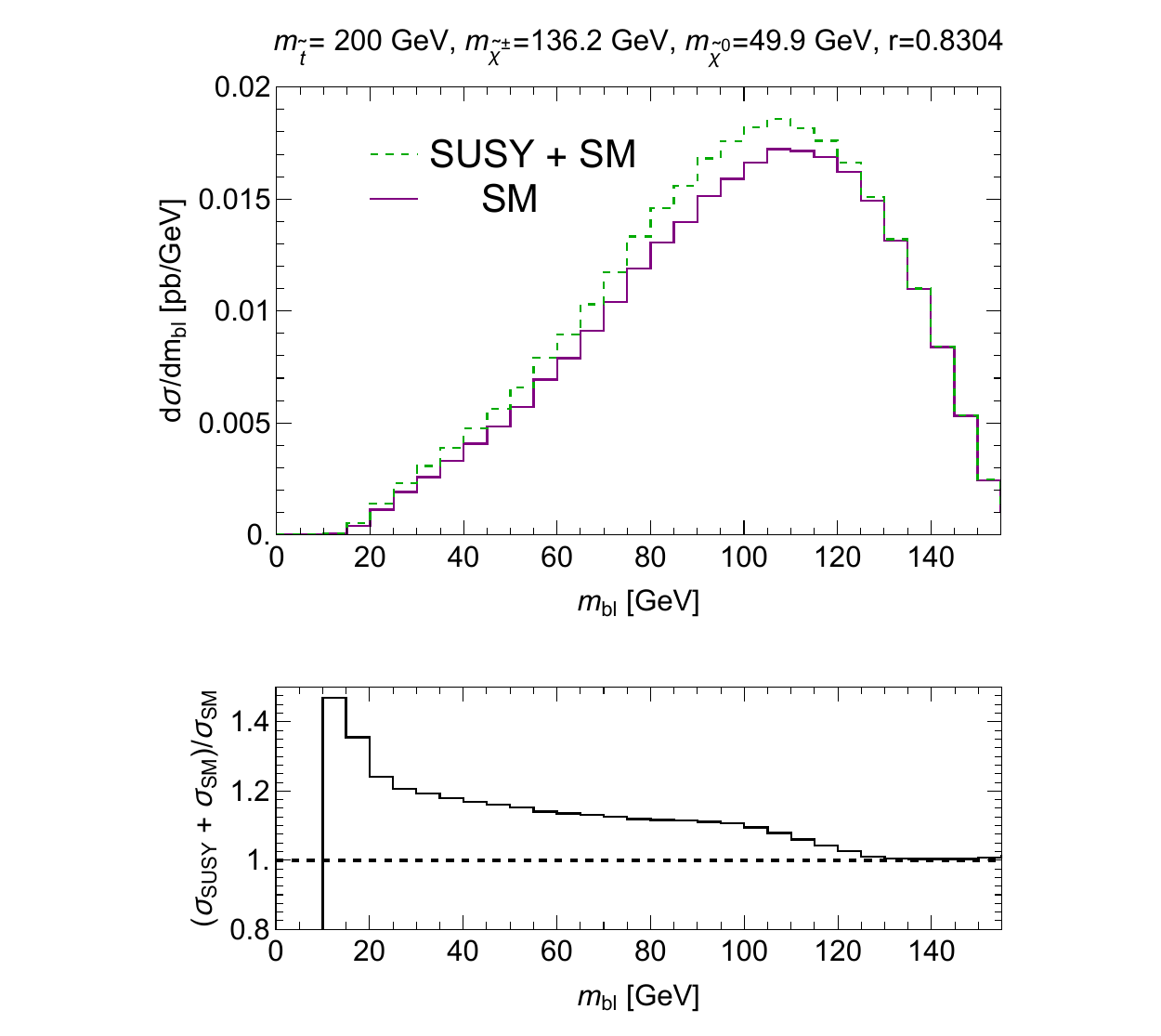} 
    \caption{BM point with signal excess}
    \label{c}
  \end{subfigure} 
  \begin{subfigure}[h]{0.5\linewidth}
    \centering
    \includegraphics[width=.7\linewidth]{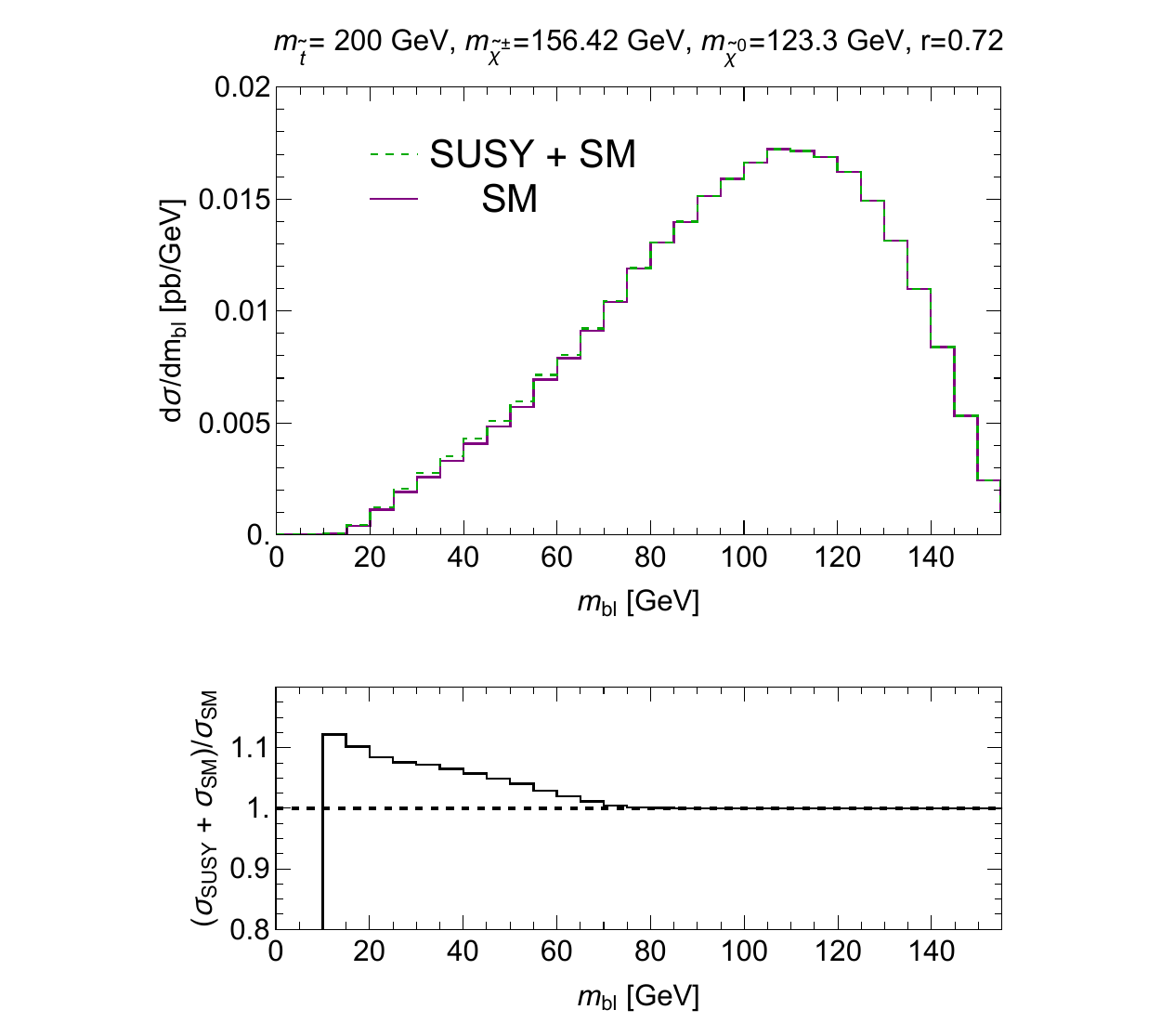}
    \caption{BM point without signal excess}
    \label{d}
  \end{subfigure}%
  \caption{}
  \label{fig:land}
\end{figure}

In order to perform a concrete study, we generate several parameter space points for the MSSM framework for three different masses of the lightest stop ($\tilde{t}_1$); $m_{\tilde{t}_1} = 180, 200, 220~$ GeV using {\tt SPheno}~4.0.3~\cite{Porod:2011nf}, interfaced with {\tt SARAH}~4.15.1~\cite{Staub:2013tta}. While generating these points, some of the input parameters are fixed such that all of these points satisfy the experimental constraints on the masses of all the \textit{sparticles} as well as the mass of the Higgs boson. While, some of the input parameters are varied to guarantee coverage of the whole parameter space for all possible masses of $\tilde{\chi}_1^{\pm}$ and $\tilde{\chi}_1^0$ for a fixed mass of $\tilde{t}_1$. However, there are several other experimental constraints, beside the ones concerning masses of BSM particles, which must be satisfied. In order to ensure that the parameter-space points satisfy all existing experimental constraints, these points are checked against searches available for recast using \SModelS~\cite{Alguero:2021dig}. \SModelS~decomposes each parameter space point into all possible topologies and compare them with every existing experimental limits and returns a value of $r$, which is a measure of the ratio of cross-section for a particular topology expected theoretically from a BSM framework to experimental predictions for that topology. If $r<1$, then that parameter space point is allowed by all existing experimental constraints. A plot showing the values of $r$ for all the parameter space points for $m_{\tilde{t}_1} = 200~$ GeV is shown in Fig.~\ref{b}. In Fig.~\ref{b}, the points in blue denotes $r<1$ and hence satisfy all constraints from experiments. 
Next, the points are simulated using {\tt Pythia}~8.3 ~\cite{Bierlich:2022pfr} using anti-$k_T$ jet algorithm~\cite{Cacciari:2008gp} and the following sets of cuts: $p_T(\ell) > 25$~GeV, $|\eta(\ell)| < 2.5$, $R(j) = 0.4$, $p_T(j) > 25$~GeV, $|\eta(j)| < 2.5$, $\Delta_R(\ell j)>0.2$, motivated by experimental analyses Ref.~\cite{ATLAS:2019onj, ATLAS:2017vgz, CMS:2016hdd}. Following the simulation, significance for each parameter space point is calculated from the $m_{b \ell}$ plot at $\sqrt{s} = 13~$ TeV and Integrated Luminosity $\mathcal{L} = 139 fb^{-1}$. While calculating the significance, we use the relative uncertainties on the background from ATLAS~\cite{ATLAS:2019onj, ATLAS:2017vgz} and CMS~\cite{CMS:2016hdd} to ensure that the error due to mismeasurement does not appear as BSM signal. Fig.~\ref{a} shows the significance of each parameter space point calculated using the relative uncertainty from ATLAS~\cite{ATLAS:2019onj, ATLAS:2017vgz} for $m_{\tilde{t}_1} = 200~$ GeV.  Therefore, parameter space points with $r<1$ (blue in Fig~\ref{b}) and high significance (orange in Fig~\ref{a}) are points of our interest which satisfy all existing experimental constraints as well as show deviation from the SM prediction in $m_{b \ell}$ distribution. The $m_{b \ell}$ distribution for one such point is shown in Fig~\ref{c}. In Fig~\ref{c} and \ref{d}, the purple curve denotes what is expected from the SM, while the green curve denotes what is obtained from a combination of a MSSM parameter space point (the signal) and the SM background. Both Fig~\ref{c} and \ref{d} exhibit a MSSM parameter space point which has $r<1$. While Fig~\ref{c} shows some signal excess in the lower regime of the $m_{b \ell}$ distribution, Fig~\ref{d} does not show any signal excess. 

\section{Conclusion}
\label{sec:con}

While presently the usual trend is to search new physics at energies beyond the current reach of the LHC but accessible by future colliders, in this article we show that a thorough study of well-known kinematic observables may hint towards the existence of New Physics. We carefully investigate the $m_{b \ell}$ distribution in the MSSM signal and the SM background with the final state: a pair of oppositely-charged dileptons + 2 $b-$ jets + $\slashed{E_T}$ and with $m_{\tilde{t}_1} \approx m_t$ for all possible values of $m_{\tilde{\chi}_1^{\pm}}$ and $m_{\tilde{\chi}_1^{0}}$. After employment of certain cuts motivated by experimental analyses ~\cite{ATLAS:2019onj, ATLAS:2017vgz, CMS:2016hdd}, the $m_{b \ell}$ distribution at $\sqrt{s} = 13~$ TeV and Integrated Luminosity $\mathcal{L} = 139 fb^{-1}$ has the potential to show some signal excess for some of the parameter space points, such as the one shown in Fig.~\ref{c}. So, to conclude, this article proposes a precise measurement and examination of well-known observables, such as $m_{b \ell}$, to search for new physics in the low energy regime before moving onto the future colliders at higher energies.  

\acknowledgments{I thank all the organisers of the $42^{nd}$ International Conference on High Energy Physics (ICHEP) 2024 for their kind hospitality. I thank my collaborators Emanuele Bagnaschi, Gennaro Corcella and Roberto Franceschini. Also thanks to Federico Meloni, Javier Montejo Berlingen, Krzysztof Rolbiecki for useful discussions. This research was supported in part by the MURPRIN2022 Grant n.202289JEW4. We thank the Galileo Galilei Institute for Theoretical Physics for the hospitality and the INFN for partial support during the completion of this work.}

\bibliographystyle{JHEP}

\bibliography{reference}

\end{document}